# Coexisting Kondo hybridization and itinerant f-electron ferromagnetism in UGe$_2$

Ioannis Giannakis[1], Divyanshi Sar[1], Joel Friedman[1], Chang-Jong Kang[2,3], Marc Janoschek[4,a], Pinaki Das[4,b], Eric D. Bauer[4], Gabriel Kotliar[2,7], and Pegor Aynajian[1]*

[1]Department of Physics, Applied Physics and Astronomy, Binghamton University, Binghamton, New York 13902, USA
[2]Department of Physics and Astronomy, Rutgers University, New Jersey 08854, USA
[3]Department of Physics, Chungnam National University, Daejeon 34134, South Korea
[4]Los Alamos National Laboratory, Los Alamos, 87545, New Mexico, USA
[5]Condensed Matter Physics and Materials Science Department, Brookhaven National Laboratory, Upton, New York 11973, USA
[a]Present address: Laboratory for Neutron and Muon Instrumentation, Paul Scherrer Institute, Villigen PSI, Switzerland
[b]Present address: Advanced Photon Source, Argonne National Laboratory, 9700 S. Cass Ave, Lemont, IL 60439

* To whom correspondence should be addressed: aynajian@binghamton.edu

**Kondo hybridization in partially filled *f*-electron systems conveys significant amount of electronic states sharply near the Fermi energy leading to various instabilities from superconductivity to exotic electronic orders. UGe$_2$ is a 5*f* heavy fermion system, where the Kondo hybridization is interrupted by the formation of two ferromagnetic phases below a 2$^{nd}$ order transition T$_c$ ~ 52 K and a crossover transition T$_x$ ~ 32 K. These two ferromagnetic phases are concomitantly related to a spin-triplet superconductivity that only emerges and persists inside the magnetically ordered phase at high pressure. The origin of the two ferromagnetic phases and how they form within a Kondo-lattice remain ambiguous. Using scanning tunneling microscopy and spectroscopy, we probe the spatial electronic states in the UGe$_2$ as a function of temperature. We find a Kondo resonance and sharp 5*f*-electron states near the chemical potential that form at high temperatures above T$_c$ in accordance with our density functional theory (DFT) + Gutzwiller calculations. As temperature is lowered below T$_c$, the resonance narrows and eventually splits below T$_x$ dumping itinerant *f*-electron spectral weight right at the Fermi energy. Our findings suggest a Stoner mechanism forming the highly polarized ferromagnetic phase below T$_x$ that itself sets the stage for the emergence of unconventional superconductivity at high pressure.**

Over the past decade, interest in exploiting ferromagnetic superconductivity with non-trivial topology dominated the field of quantum matter due to their robust functionalities in quantum information[1–3]. Yet, quantum materials that exhibit natural coexistence of ferromagnetism and superconductivity remain rare. To date, only a handful of such synthesized single crystalline materials exist, with the majority being uranium-based heavy fermion compounds including $UGe_2$[4], $URhGe$[5–7], $UCoGe$[8] and the recently discovered $UTe_2$[9]. In these compounds, the $5f$-electrons play a critical role in the emergence of the exotic superconductivity[10], making it particularly crucial to understand their controversial normal state behavior.

$f$-electrons in heavy fermion compounds exhibit dual characteristics of being itinerant and localized, driving an electronic competition between magnetism, very often of antiferromagnetic character, and heavy Fermi liquid behavior with quenched magnetic moments[11–15]. Recent experimental and computational work demonstrated the dual nature of $5f$ electrons in $USb2$[16,17], an antiferromagnetic heavy fermion system, through orbital selectivity, providing a natural explanation of how localized magnetism and itinerant heavy fermions of the same uranium $5f$ electrons coexist. The emergence of f-electron ferromagnetism and its interplay with Kondo coherence remains much less explored.

The ferromagnetic heavy fermion $UGe_2$ displays an interesting phase diagram[5,18,19]. At ambient pressure, a second order paramagnetic-to-ferromagnetic phase transition at a relatively high $T_c \sim$ 52 K [18] is followed by a crossover meta-magnetic transition from a weakly polarized ferromagnetic state, FM1, to a strongly polarized ferromagnetic state FM2 [20] at $T_x \sim$ 32 K. Transport measurements show the emergence of the Kondo-lattice effect at $T_K \sim$ 110 K, well above $T_c$[19]. How the Kondo effect is impacted by the emergent ferromagnetism at and below $T_c$ remains a question to be answered. Below $T_c$, specific heat measurements display a broad hump centered around $T_x$ [19,21] which, along with magnetization measurements [19,22] and neutron scattering[23] shows the presence of itinerant and localized subset of the uranium $5f$ electrons. In the same temperature range, Hall effect studies reveal a rapid increase of charge carriers below $T_x$ suggestive of some sort of Fermi surface reconstruction [24]. This reconstruction is argued to be caused by the sudden delocalization of the uranium $5f$ electrons. The microscopic origins of $T_c$ and $T_x$ are particularly important for the mechanism of emergent exotic superconductivity in $UGe_2$. With the application of hydrostatic pressure, the pressure-dependent FM2 transition line $T_x(P)$ decreases and terminates at the maximum of the emergent superconducting dome at $P_x \sim$ 1.2 GPa, suggesting its fluctuations and destruction are directly related to the mechanism of superconductivity[7]. Furthermore, the superconducting dome only persists inside the FM1 phase, where both phases simultaneously disappear at the exact same pressure $P_c \sim$ 1.5 GPa indicating an intimate relation between the ferromagnetism and superconductivity[25,26]. Therefore, the emergence of itinerant $5f$-electrons through the Kondo effect in $UGe_2$ and their evolution across $T_c$ and $T_x$ forms the low temperature normal state near the Fermi energy ($E_F$) out of which ferromagnetic superconductivity develops.

Theoretical understanding of the nature of the ferromagnetic state below $T_x$ is controversial. One mechanism comes from the phenomenological ideas following the rigid-band Stoner approach, where two sufficiently sharp and narrowly separated density-of-state peaks located near the Fermi energy form the majority and minority spin bands[27]. Another idea involves charge and spin density waves emerging below $T_x$[18]. In either case, direct experimental signature of a double peak structure or density waves have not been observed to date.

A sharp resonance in the density of states can naturally arise in heavy fermion compounds, whose energy relative to $E_F$ depends on the valence of the *f*-electrons in the material system[28,29]. Scanning tunneling microscopy (STM) has the spatial and energy resolution to probe the sharp resonance and it's possible splitting[30]. Yet, due to the lack of a natural cleaving plane in $UGe_2$, which is crucial to obtain clean surfaces for STM, such an experiment have so far not been carried out. Here we use STM to probe the local electronic states and their temperature evolution near $E_F$ in single crystal $UGe_2$. We find multiple peaks in the density of states located near the chemical potential above $T_c$. The finding is in qualitative agreement with our DFT+ Gutzwiller calculations presented here, attributing their origin to the different 5*f*-electronic orbital characters. With lowering of temperature, the two peaks located nearest to $E_F$ strengthen and narrow. Below $T_c$ and particularly near $T_x$ ~ 32 K, the peaks split, forming additional kinks that further develop and evolve with temperature suggesting a Stoner mechanism of the ferromagnetic order. Our finding indicates a significant degree of itinerant character of the *f*-electrons through Kondo hybridization and the itinerant nature of the ferromagnetism involving the same *f*-electrons. At the lowest temperature, a sharp *f*-electronic density of states is formed at $E_F$ setting the stage for the emergence of ferromagnetic spin-polarized superconductivity at higher pressure.

Figure 1a, b shows STM topographs of the (010) surface of single crystal $UGe_2$, in-situ cleaved in our ultra-high vacuum, variable temperature STM. Cleaving exposes alternating terraces of two chemically different surfaces, termed A and B in Fig.1c. While surface A displays spatial uniformity with no atomic corrugation indicating the extended nature of the electronic states, surface B undergoes a surface reconstruction, whose quasiperiodic structure differs between different cleaves, as seen in Fig.1a, b. The asymmetry in the step height (A→B > B→A, with A↔A = B↔B ≡ b-axis unit cell) between the different terraces allows us to compare the results to the crystal structure. Assuming only a single chemical bond breaking during the cleaving process leads us to identify surfaces A and B as uranium and germanium terminated, respectively (Fig.1c). Such an assumption is justified by the fact that only two surfaces have been observed based on ten different sample cleaves. The Ge surface with two Ge atoms per ac-plane unit cell, as compared to a single U or Ge atom per ac-plane for all other layers (see Fig.1), is also more likely to undergo surface relaxation and buckling. Nevertheless, the surface assignment here does not change the conclusions reached below.

STM spectra probed on the two surfaces just above $T_c$ reveal two asymmetric low-energy peaks in the density of states in close proximity to the chemical potential, whose intensities are different on the two surfaces (Fig.1d). An additional broader high-energy peak near ~100 meV is also observed. Note that the intensity of the peaks on surface-B are spatially non-uniform and is further elaborated below. At low temperatures (8 K), the low energy peaks undergo a splitting with the emergence of a kink/shoulder (Fig.1e). The high-energy peak remains mostly unchanged.

Peaks in the density of states near $E_F$ is the hallmark of itinerant *f*-electron systems and have been seen in previous STM experiments in various 4*f* and 5*f* heavy fermions[31]. They are the results of the Kondo hybridization of localized f-orbitals with conduction electrons. To corroborate this observation, we compute the electronic structure of $UGe_2$ by employing the generalized gradient approximation to density functional theory (DFT) in combination with the Gutzwiller approximation (DFT+Gutzwiller)[32]. This method captures electronic correlations beyond the single-particle picture of DFT and has been successfully applied to other *f*-electron systems such as $UO_2$[33]. The local Coulomb interaction strength and the Hund's coupling constant are U = 6.0 eV and J = 0.57 eV, respectively, for the correlated U 5*f*-orbital. The DFT+Gutzwiller calculations were performed at T = 0 K within the paramagnetic phase. Within DFT there is a significant mixing between U $5f_{5/2}$ and $5f_{7/2}$ states, so their spin-orbit splitting is not apparent. X-ray photoemission spectroscopy[34] and x-ray magnetic circular dichroism[35] measurements indicate that there is a clear spin-orbit splitting between U $5f_{5/2}$ and $5f_{7/2}$. Furthermore, the $5f_{5/2}$ level lies below the $5f_{7/2}$ level and the magnitude of the splitting is around 1.1 eV. This feature is well captured by the DFT+Gutzwiller calculations, which gives a spin-orbit splitting of ~1.5 eV. Our calculations indeed show multiple uranium 5*f*-electron peaks with different orbital character to reside near $E_F$ (Fig.1f). More specifically, three major peaks located at energies of -18meV, +35meV and +66meV that have characters of U (J = 5/2, $m_J$ = ±1/2, ±5/2, ±1/2), respectively, are qualitatively consistent with the high temperature experimental data observed on surface A and/or B at ~ -20meV, ~ +25meV, and ~ +100 meV (Fig.1d, e).

While the relative widths and weights of the spectral lineshapes are spatially uniform on surface A (see Fig.2a), they vary significantly on surface B due to the structural inhomogeneity induced by surface reconstruction, as seen in Fig.2b. For example, looking at the different peaks, one can see that their intensity can be dramatically suppressed depending on the spatial location on the surface. This also applies to the spectra in Fig.1d, where at high temperature only the negative peak is observed with almost no peak intensity on the positive side. The disappearance of the positive bias peaks is due to the particular location of the spectrum on the surface and other locations (not shown here) do reveal a finite peak at positive and negative biases. This spatial variation renders studying the detailed temperature dependent evolution unreliable on surface B. We therefore focus on surface A (U surface) to probe the temperature dependence of the spectra across the two ferromagnetic transitions. Figure 3 shows our temperature dependent spectroscopy measurements carried out on surface A. The dI/dV conductance were measured in a constant current mode, $I_{set}$, and a bias voltage applied to the sample, $V_{bias}$, with a bias modulation of $V_{mod}$. Two sets of data with different experimental settings (energy-resolution) of $I_{set}$ = 150 pA, $V_{bias}$ = 500 meV, $V_{mod}$ = 5 meV (Figure 3a) and $I_{set}$ = 1 nA, $V_{bias}$ = 200 meV, $V_{mod}$ = 1 meV (Figure 3b) are shown. The spectra display an asymmetric resonance analogous to that seen in other heavy fermion systems[31,36–41]. The observed resonance is the manifestation of Kondo hybridization, delocalizing the *f*-electrons and merging them into the Fermi sea starting already at temperatures above $T_c$. As temperature is lowered below $T_c$, we observe the sharp kink at $E_1$ (near the Fermi energy) starting to develop particularly below ~35 K (see insets in Fig.3a, b). At the lowest measured temperature of 8 K, a clear double peak structure can be resolved with a peak separation ($E_1 - E_2$) of ~ 16 meV.

The strong temperature broadening of the spectral lineshapes makes it difficult to pin-point the onset of the $E_1$ kink in the raw data. In Fig.3c, d, we show the 2$^{nd}$ derivative of the spectral lineshapes, which highlights the kink-structure near the Fermi energy. While the temperature evolution is weak above ~ 35 K, it becomes more pronounced at lower temperatures. To better visualize this behavior, we contrast in Fig.3e, the high resolution spectra with a model Fano lineshape. A Fano lineshape in STM spectra resembles an asymmetric resonance peak due to interference between the two tunneling paths from the tip to the heavy (resonance) and light (continuum) electronic states of a Kondo lattice and has been widely used in STM analysis of heavy fermion systems[42–46]. The equation below represents the Fano lineshape

$$\frac{dI}{dV} \propto A \frac{(\frac{V-E}{\Gamma}+q)^2}{1+(\frac{V-E}{\Gamma})^2}$$

where E characterizes the resonance energy, $\Gamma$ the resonance linewidth expressed as Half Width at Half Maximum (HWHM) and *q* is the tip-sample coupling, also known as the asymmetry parameter. A is related to the amplitude of the resonance. Figure 3e shows the data and the corresponding fit to a single Fano lineshape. We observe at 55K (T>$T_c$) that the data can be nicely modeled by a single resonance. At 35 K however, we can see that the data deviates from a single Fano lineshape particularly in the energy range of ±10 meV, where a second resonance develops and grows stronger with further cooling. We therefore use a two Fano lineshape model (one centered at $E_1$ and another at $E_2$) to fit the temperature dependent data. Figure 4a shows the data together with their corresponding fit to the summation of two Fano resonances. For all temperatures, the model fit shows an excellent agreement with the data. No additional background is used in the model. The extracted resonance amplitude, widths, and energies are displayed in Fig.4 b, c, d respectively.

Looking at the amplitude of the resonances (Fig.4b), we first note that both resonances ($E_1$ and $E_2$) weaken with increasing temperature. While the $E_2$ resonance amplitude remains finite and large with no apparent anomaly at $T_c$, the $E_1$ resonance fades and within the experimental resolution becomes negligible above 35 K, as is reflected by the diverging error bars, which renders their values meaningless above 35 K. The extracted linewidths of the resonances also paint a similar picture. At the lowest temperature, the linewidths of the $E_1$ and $E_2$ resonances saturate at values of ~ 7meV and ~13meV, respectively. With increasing temperature, the $E_2$ resonance increases and within the experimental resolution follows the conventional temperature dependence expected in Kondo lattice systems $\Gamma(T) = \sqrt{(\pi k_B T)^2 + 2(k_B T_K)^2}$ [38,39]. Plotting $\Gamma(T)$ for a $T_K$ of 110 K extracted from transport measurements (blue line in Fig.4c) reveals a good agreement with the experimental data. On the other hand, the $E_1$ linewidths grow rapidly and diverge above 35 K, where the error bars span the entire y-range and the data are therefore omitted for T > 35 K from Fig.4c. The rapid decrease of spectral weight of the $E_1$ resonance together with its diverging linewidths with increasing temperature makes it difficult to ascertain its high-temperature evolution, particularly above 35 K. The extracted energies below 35 K show no significant temperature dependence.

We now turn to identify the origin of the observed double-peak structure at low temperatures. One possibility is the indirect Kondo hybridization gap[47]. However, this scenario can be discarded in UGe$_2$ for two reasons. First, the observed band splitting occurs far below the Kondo lattice temperature of 110 K extracted from transport measurements. In fact, looking at the temperature dependence of the $E_1$ linewidth, one can see that it deviates dramatically from thermal broadening and $\Gamma(T)$ (red line in Fig.4c) and diverges near 35 K. Second, contrasting our observation in UGe$_2$ with the Kondo resonance in antiferromagnetic USb$_2$[16], non-magnetic UTe$_2$[9], and URu$_2$Si$_2$[38,40] above its hidden order temperature, we see that all these uranium-based heavy fermion systems display a single Fano resonance above the chemical potential. Yet, none show a splitting or an indirect Kondo hybridization gap opening at low temperature, regardless of their magnetic or Kondo-coherence temperatures. Therefore, the splitting that we observe is likely not due to an indirect hybridization gap, which should show up similarly in these other U-based systems as well. We therefore turn to ferromagnetism as a possible origin of the band splitting (~ 16 meV) that we observe.

In itinerant ferromagnets, below their magnetic transition, the spin-majority and minority bands split due to the ferromagnetic exchange. Our observation is consistent with this Stoner mechanics of itinerant ferromagnetism. Indeed, the observation of a Kondo resonance above $T_c$ and its further evolution below $T_c$ provides spectroscopic evidence of the itinerant character of the *f*-electrons and therefore of the ferromagnetism in UGe$_2$. This agrees with the fact that the ordered moment in the ferromagnetic phase is 1.4 $\mu_B$/U, much smaller than the effective paramagnetic moment of 2.7 $\mu_B$/U[19]. Evidence of itinerant ferromagnetism and band splitting is also seen in the optical spectroscopy of UGe$_2$ below $T_c$, where low energy excitations with an energy of E ~ 13.6 meV[48] have been observed. The latter is comparable to the separation of the $E_1$-$E_2$ resonances seen here (Fig.4d). Similarly, a Stoner gap of the order of 40 K has also been inferred from magnetic neutron diffraction[49].

Overall, our data reveal a Kondo resonance with a characteristic Kondo-lattice temperature of 110 K, consistent with transport measurements[19], that survives in the ferromagnetic phase. In heavy fermion systems, the magnetic ground state is generally in competition with the Kondo quenching of the magnetic moments that lead to the famous Doniach diagram. This holds true in Ce-based heavy fermions as seen in CeRhIn$_5$ and CeCoIn$_5$[50,51]. In UGe$_2$ however, the two phenomena seem to coexist with no apparent competition, as seen by the largely unaffected Kondo resonance crossing $T_c$. A similar conclusion was reached in USb$_2$[16,17]. Gradually below ~35 K, an additional kink/shoulder ($E_1$ resonance) develops that coincides with the highly polarized FM2 phase that emerges as a crossover below $T_x \sim 32$ K. The $E_1$ resonance shifts the *f*-spectral weight closer to $E_F$, which leads to an enhanced electronic effective mass. The latter has been observed in the optical spectroscopy measurements[48]. Hall measurements are also consistent with this picture, where a rapid increase of charge carriers indicating some sort of Fermi surface reconstruction is observed below $T_x$ [24]. Altogether, these observations indicate the splitting of the resonance that likely onsets at $T_c$ and becomes more pronounced below the crossover temperature $T_x$ following the rigid band Stoner model[27,49], leading to a spin polarized *f*-electron band ($E_1$) at the Fermi energy, which sets the stage for a spin polarized superconductivity to emerges at high pressure. Future spin polarized STM study on the relative peak intensities of $E_1$ and $E_2$ will confirm the findings here, but is beyond the scope of this study.

**Acknowledgments**

Work at Binghamton University is supported by the U.S. National Science Foundation (NSF) CAREER under award No. DMR-1654482. Work at Los Alamos National Laboratory was performed under the auspices of the US Department of Energy, Office of Basic Energy Sciences, Division of Materials Sciences and Engineering.

**Author Contributions:**

I.G. performed the STM measurements. I.G. and P.A. performed STM data analysis with help from D.S. and J.F.. C-J.K. and G.K. performed DFT + Gutzwiller calculations. M.J., P.D. and E.B. synthesized and characterized the materials. P.A. wrote the manuscript.

Figure captions:

Figure 1

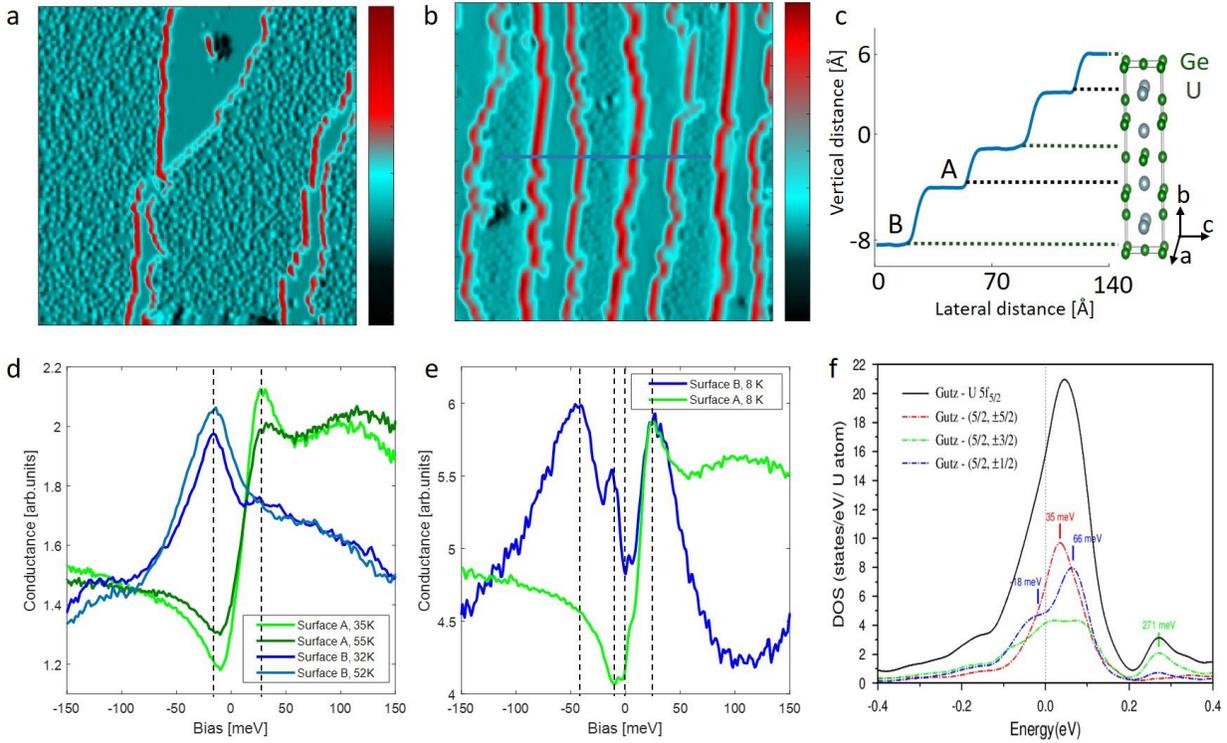

Figure 1: (a, b)Topographic image of cleaved UGe$_2$ revealing subatomic terraces with alternating chemically different surface terminations, one of which undergoes a surface reconstruction. (c) Comparison of the terrace heights extracted from the blue line-cut in (b) with the crystal structure of UGe$_2$ suggesting the breaking of a single bond (U-Ge2) that exposes the reconstructed Ge2 surface (B) and U surface (A). (d, e) STM dI/dV spectra on the two surfaces at high (d) and low (e) temperatures. Note that the data in (d, e) are not taken on the same spatial location. (f) DFT+Gutzwiller calculation of the electronic density of states in UGe$_2$ in the non-magnetic phase. Note that U 5f$_{7/2}$ due to spin-orbit coupling are at much higher energies (1 – 2 eV).

Figure 2

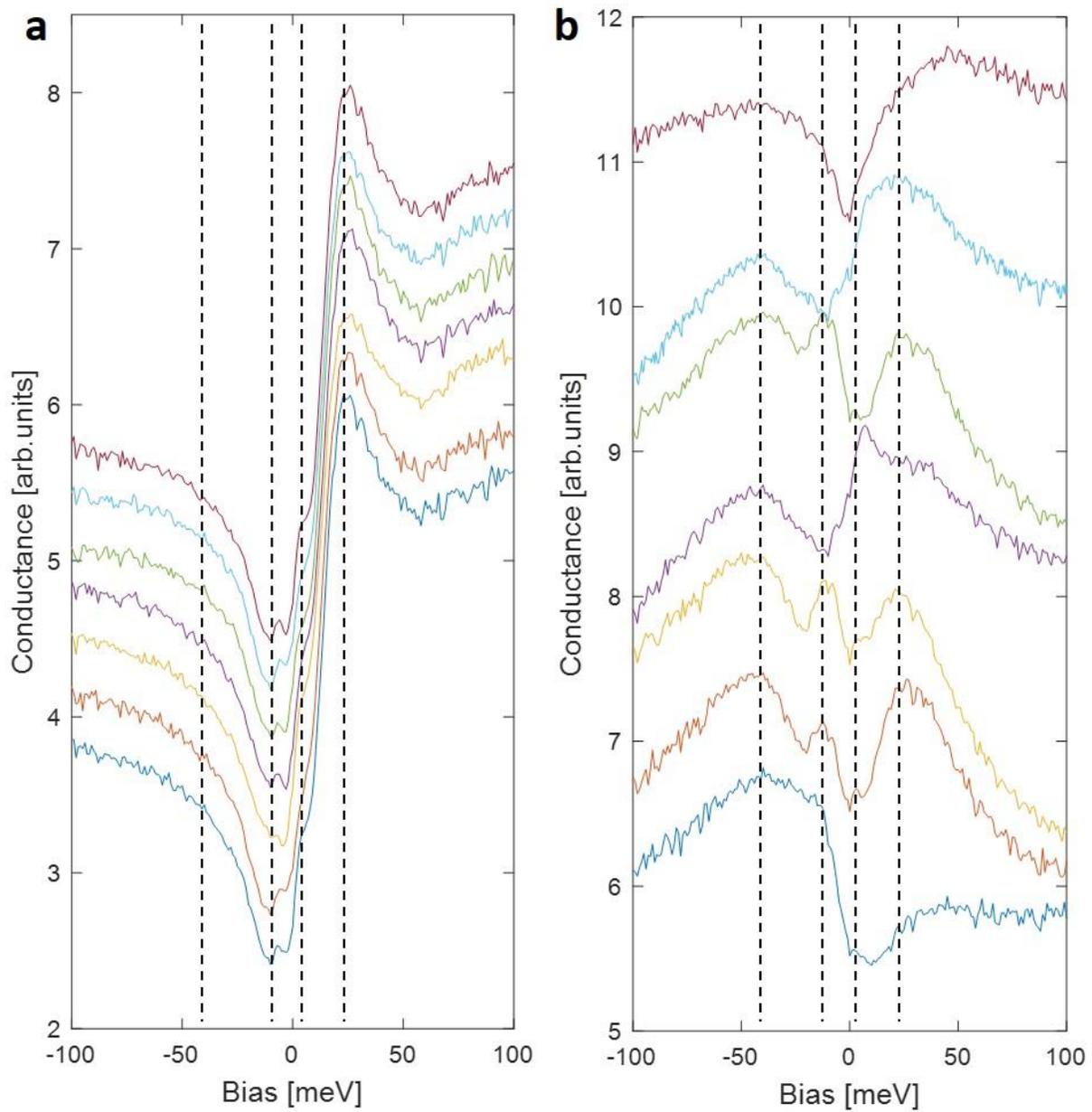

*Figure 2: STM spectra taken at different locations on surfaces A (a) and B (b) at 8 K. The spectra in (a) reveal spatial uniformity, whereas those in (b) are spatially inhomogeneous due to the surface inhomogeneity. The lines correspond to the energies of the observed peaks that are similar on both surfaces.*

Figure 3

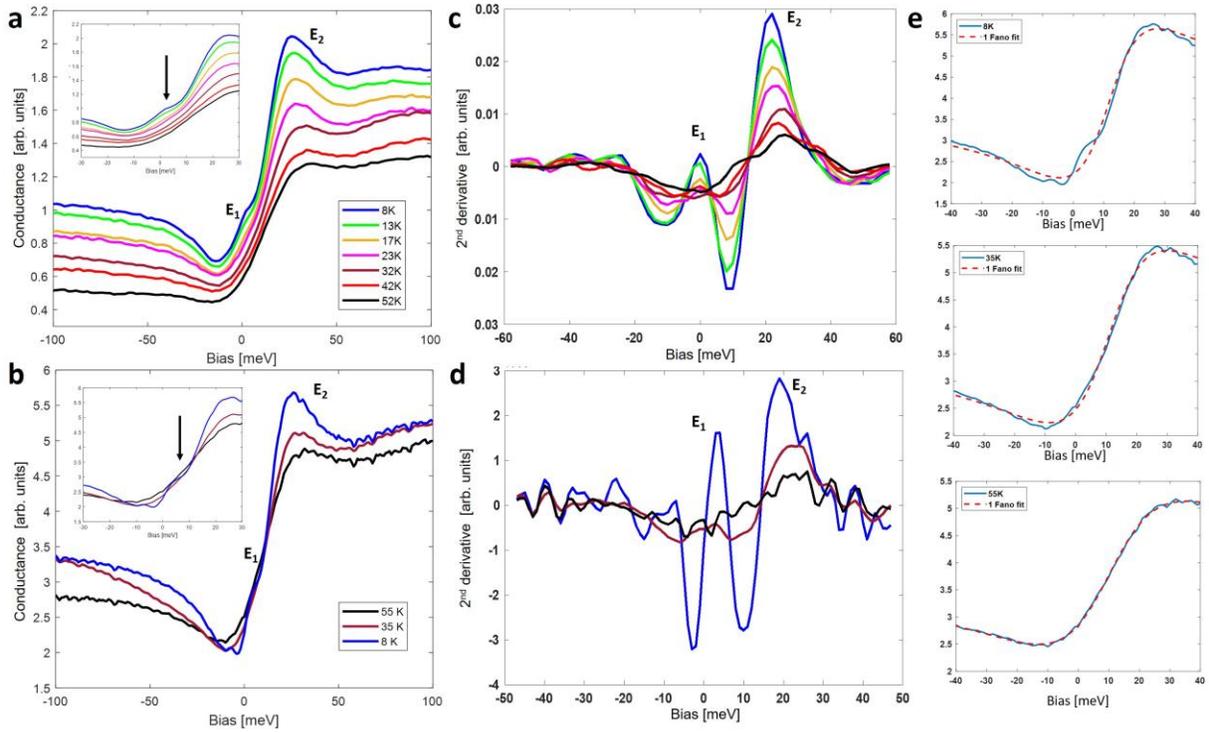

*Figure 3: (a, b) STM spectroscopy on surface-A as a function of temperature with two different experimental settings described in the text. The spectra reveal a Fano lineshape ($E_2$ resonance) located above the chemical potential. As temperature is lowered below $T_c$, a second resonance emerges near the Fermi energy ($E_1$ resonance) below ~ 35 K. (c, d) Second derivative of the spectra in (a, b) showing the evolution of the $E_1$ and $E_2$ resonances as a function of temperature. (e) Fitting of the data in (b) to a single Fano lineshape. The spectrum above $T_c$ can be explained by a single resonance, while at lower temperatures, the data deviates from a single Fano lineshape.*

Figure 4

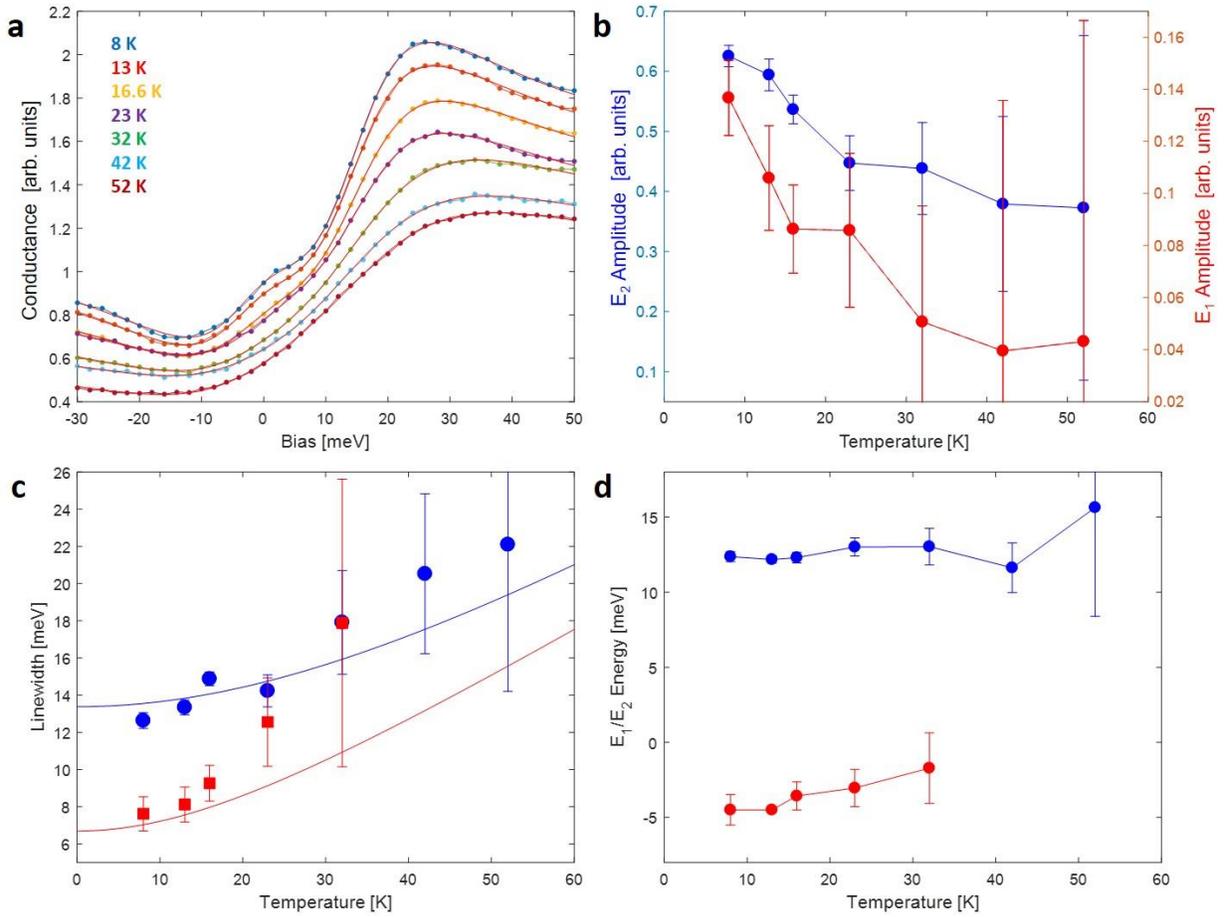

Figure 4: (a) STM spectroscopy and corresponding two Fano-lineshape model fit at different temperatures. (b,c,d) extracted Fano amplitude, width and energy as a function of temperature. Red data points correspond to the $E_1$ resonance near $E_F$ whereas the blue data points correspond to the $E_2$ resonance above the chemical potential. The lines in (c) represent $\Gamma(T)$ for a TK of 110K (blue) and TK of 55K (red).